\documentclass[prb,showpacs,twocolumn,aps]{revtex4}
\usepackage[dvipdfm]{graphicx}
\usepackage[dvipdfm]{color}
\usepackage{dcolumn}
\usepackage{amsmath}
\usepackage{amssymb}

\newcommand{\lco}{$\rm La_{2}CuO_4$}
\newcommand{\lsco}{$\rm La_{2-{\textit x}}Sr_\textit{x}CuO_4$}

\newcommand{\lsoco}{$\rm La_{1.99}Sr_{0.01}CuO_4$}

\newcommand{\lbco}{$\rm La_{2-{\it x}}Ba_{\it x}CuO_4$}
\newcommand{\lbcoe}{$\rm La_{1.875}Ba_{0.125}CuO_4$}

\newcommand{\leco}{$\rm La_{1.8}Eu_{0.2}CuO_4$}

\newcommand{\lesco}{$\rm La_{1.8- \textit{x}}Eu_{0.2}Sr_\textit{x}CuO_4$}

\newcommand{\lescoex}{$\rm La_{1.79}Eu_{0.2}Sr_{0.01}CuO_4$}

\newcommand{\lnco}{$\rm La_{1.7}Nd_{0.3}CuO_4$}

\newcommand{\lndco}{$\rm La_{2- \textit{y}}Nd_\textit{y}CuO_4$}

\newcommand{\sus}{susceptibility}

\newcommand{\DM}{Dzyaloshinsky--Moriya}

\newcommand{\pla}{$\rm CuO_2$}
\newcommand{\oct}{$\rm CuO_6$}

\begin{document}

\title{Electronic interlayer coupling in the LTT phase of $\bf La_{1.79}Eu_{0.2}Sr_{0.01}CuO_4$}
\author{M. H\"ucker}
\affiliation{Brookhaven National Laboratory, Upton, New York 11973-5000, USA}

\date{\today}

\begin{abstract}
The electronic interlayer transport of the lightly doped antiferromagnet $\rm
La_{1.79}Eu_{0.2}Sr_{0.01}CuO_4$ has been studied by means of magneto-resistance
measurements. The central problem addressed concerns the differences between the
electronic interlayer coupling in the tetragonal low-temperature (LTT) phase and the
orthorhombic low-temperature (LTO) phase. The key observation is that the spin-flip
induced drop in the $c$-axis magneto-resistance of the LTO phase, which is characteristic
for pure \lsco , dramatically decreases in the LTT phase. The results show that the
transition from orthorhombic to tetragonal symmetry and from collinear to non-collinear
antiferromagnetic spin structure eliminates the strain dependent anisotropic interlayer
hopping as well as the concomitant spin-valve type transport channel. Implications for
the stripe ordered LTT phase of \lbco\ are briefly discussed.
\end{abstract}

\pacs{74.72.Dn, 74.25.Fy, 74.25.Ha, 61.50.Ks}

\maketitle

\section{Introduction}

Due to their layered structure high-$\rm T_c$ superconductors such as \lsco\ have
strongly anisotropic properties. The electronic conductivity perpendicular to the \pla\
planes is between two and four orders smaller than in the planes, and the effective
interlayer superexchange about five orders weaker than the Cu-O-Cu in-plane
superexchange.~\cite{Preyer89,Thio88,Thio90} Nevertheless, a finite electronic interlayer
coupling is essential for 3D antiferromagnetic (AF) order or 3D bulk superconductivity
(SC) to occur.~\cite{Mermin66}
\lco\ has been an ideal playground for experimental and theoretical studies of interlayer
interactions.~\cite{Thio88,Gozar04a,Huecker04c,Shekhtman94a,Viertio94a,Benfatto06a} It is
amenable to doping and offers examples where a small modification of the crystal
structure can change the ground state. Particularly interesting is the case of \lbco\
with $x \simeq 1/8$, where bulk SC is suppressed and replaced by a static order of charge
and spin stripes.~\cite{Moodenbaugh88a,Maeno91,Tranquada95a,Fujita04a} Concomitant with
stripe order a transition from LTO to LTT is observed.~\cite{Axe89} There is growing
evidence that the stripe ordered LTT phase causes an electronic decoupling of the \pla\
planes.~\cite{Li07c,Berg07a,Chubukov07a,Tranquada08a,Ding08a} The complexity of the
involved electronic, magnetic, and structural interactions, however, poses a challenge
for an unambiguous experimental analysis.

Therefore, the focus of the present work lies on lightly doped samples ($x<0.02$), where
the influence of structure and magnetism on the electronic transport may be deciphered
more easily. There is no SC or long range stripe order involved, and the AF order is
commensurate as long as one does not cool below the spin glass
transition.~\cite{Matsuda02a} Early magneto-resistance and magnetization measurements on
\lco , and more recently on \lsoco\ have shown that in the LTO phase the electronic
interlayer transport depends on how the AF sublattices are stacked along the
c-axis.~\cite{Thio88,Thio94,Shekhtman94a,Ando03a} Here, similar magneto-resistance
experiments on a \lescoex\ single crystal are reported. This compound exhibits the same
sequence of structural transitions as \lbcoe , thus providing the opportunity to analyze
the electronic interlayer coupling in the lightly doped LTT phase.

The paper is organized as follows. In the next section the experimental methods are
described. The results are presented in Sec.~\ref{results}. There are three parts with
focus on the crystal structure, the magneto-transport, and complementary magnetization
measurements. In Sec.~\ref{discussion} it is shown how these properties are connected and
enable an interpretation of the electronic interlayer transport in the LTT phase. At the
end of this section implications for \lbco\ are pointed out.

\section{Experiment}
\label{exp}
The \lescoex\ single crystal with a diameter of $\varnothing$ 5~mm was grown by the
travelling-solvent floating-zone method in an atmosphere of flowing oxygen gas at a
pressure of $p({\rm O_2})=5$~atm. As grown the crystal contains a considerable amount of
excess oxygen, which was removed by annealing in Ar at 900$^\circ$C for 24~h.
The electric resistance $\rho$ of bar shaped samples was measured with the four terminal
method for currents $I$ and magnetic fields $H$ applied perpendicular and parallel to the
\pla\ planes. The leads were attached with silver epoxy, carefully cured to reduce the
contact resistance.

The x-ray diffraction experiments were performed at beamline X22C of the National
Synchrotron Light Source at a photon energy of 8.9~keV. Scattering vectors ${\bf
Q}=(h,k,\ell)$ are specified in units of $(2\pi/a, 2\pi/b, 2\pi/c)$, where $a$, $b$ and
$c$ are the lattice parameters of the orthorhombic unit cell.~\cite{Huecker06a} At room
temperature $a=5.35$~$\rm \AA$, $b=5.42$~$\rm \AA$, and $c=13.05 $~$\rm \AA$, while at
20~K in the LTT phase $a=b=5.38$~$\rm \AA$, and $c=13.0 $~$\rm \AA$. The experiment was
performed in reflection geometry on a polished surface which, due to twinning in the
orthorhombic phase, is normal to either [1,~0,~0] or [0,~1,~0].

The static magnetization $M(H)$ at constant temperatures and the static \sus\
$\chi(T)=M(T)/H$ at constant magnetic fields were measured with a SQUID (superconducting
quantum interference device) magnetometer.
All studied crystal pieces are twinned in the $ab$-plane, i.e., when measuring along the
orthorhombic in-plane axes one averages over domains with the $a$ and $b$ axes
interchanged. The degree of twinning was determined for each sample by bulk magnetization
measurements and will be indicated wherever of relevance. Data with dominant contribution
of the $a$-axis ($b$-axis) will be indexed with $a^+$ ($b^+$).

\section{Results}
\label{results}
\subsection{Crystal structure}
\label{xrd}
\begin{figure}[t]
\center{\includegraphics[width=0.7\columnwidth,angle=0,clip]{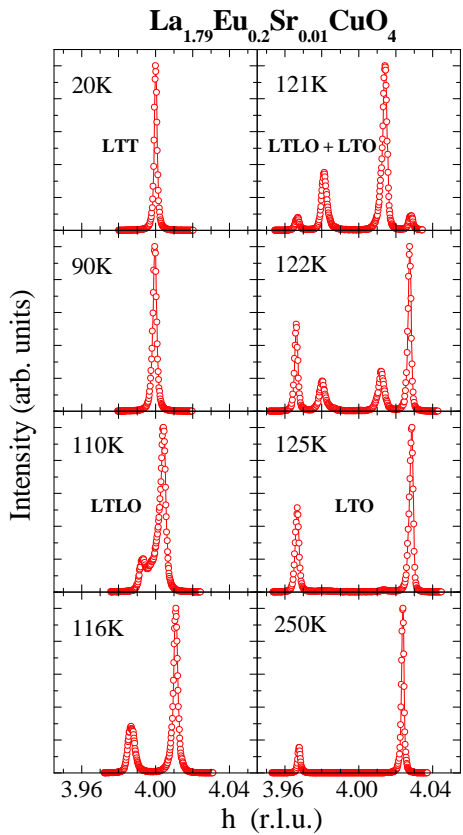}} \caption[]{(color
online) Scans through the (4,~0,~0)/(0,~4,~0) Bragg reflections of the two twin domains
of the crystal at different temperatures.} \label{fig1}
\end{figure}
\begin{figure}[t]
\center{\includegraphics[width=0.8\columnwidth,angle=0,clip]{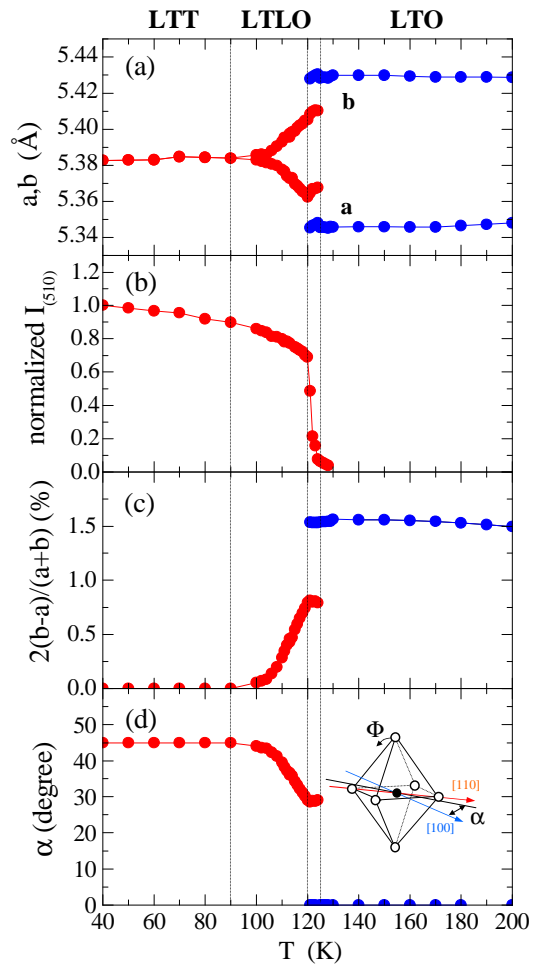}}
\caption[]{(color online) Structural transition
LTO$\leftrightarrow$LTLO$\leftrightarrow$LTT. (a) Lattice parameters $a$ and $b$. (b) Normalized
sum of the integrated intensities of the (5,~1,~0) and (-1,~5,~0) twin domain reflections. (c)
Orthorhombic strain $2(b-a)/(a+b)$ in percent of the average in-plane lattice constant.
(d) Calculated in-plane rotation $\alpha$ of the octahedral tilt axis with respect to its
direction, [1,~0,~0], in the LTO phase. See inset.} \label{fig2}
\end{figure}

Single crystal x-ray diffraction experiments were performed since the interpretation of
the transport measurements requires a detailed knowledge of the structure. At high
temperature~\cite{HTT} \lescoex\ transforms from the high-temperature tetragonal (HTT)
phase with space group $I4/mmm$ to the LTO phase with space group $Bmab$. This transition
also occurs in \lsco .~\cite{Boeni88,Radaelli94a} However, the Eu-doped compound shows a
second transition at $T_{\rm LT}$ from LTO to LTT with space group $P4_2/ncm$. The nature
of these transitions has been discussed in numerous
studies.~\cite{Axe89,Crawford91,Buechner94c,Simovic03a,Huecker06a} In first approximation
all phases can be described by different pattern of tilted \oct\ octahedra, parameterized
by the tilt angle $\Phi$ and the tilt direction $\alpha$, measured as the in-plane angle
between the tilt axis and the [100] direction, see Fig.~\ref{fig2}(d). In the HTT phase
$\Phi=0^\circ$. In the LTO phase $\Phi>0^\circ$ and $\alpha = 0^\circ$, while in the LTT
phase $\Phi>0^\circ$ and $\alpha = 45^\circ$. $\Phi$ is on the order of several degree
and approximately the same in the LTO and LTT phase. Thus, the major change at the
LTO$\rightarrow$LTT transition is a $45^\circ$ rotation of the tilt axis. Note that in
the LTT phase $\alpha$ changes sign from plane to plane, i.e., the tilt axes in adjacent
layers are orthogonal.
There have been questions about whether lightly doped \lesco\ becomes truly tetragonal,
or assumes the low-temperature less-orthorhombic (LTLO) phase with space-group $Pccn$,
which is an intermediate phase between LTO and LTT with $0^\circ < \alpha <
45^\circ$.~\cite{Buechner94b,Keimer93,Crawford91} The following results will clarify this
point.

Figure~\ref{fig1} shows scans through the (4,~0,~0)/(0,~4,~0) reflections. Above 125~K
their is only one pair of reflections, i.e., the sample is in the LTO phase. Upon cooling
two additional peaks with reduced split appear, indicating a coexistence of the LTO and
the LTLO phase. Below 120~K the transformation towards LTLO is completed. The
orthorhombic strain quickly decreases and below 90~K the crystal is in the LTT phase. A
summary of the temperature dependence of some structural properties is given in
Fig~\ref{fig2}. Panel (a) shows the lattice parameters $a$ and $b$, panel (b) the sum of
the integrated intensity of the (5,~1,~0)/(-1,~5,~0) super structure reflections which
are allowed in the LTLO and LTT phases only. Figure~\ref{fig2}(c) shows the orthorhombic
strain $2(b-a)/(a+b)$, and Fig.~\ref{fig2}(d) calculated values for $\alpha = 0.5 \cdot
{\rm acos}[(b-a)/(b_0-a_0)]$, where $b_0$ and $a_0$ are the lattice parameters in the LTO
phase just above the structural transition. In the LTO phase $\alpha$ was set zero. The
x-ray diffraction results clearly demonstrate that the low temperature transition in
\lescoex\ is a sequence of two transitions: a discontinuous LTO$\leftrightarrow$LTLO
transition and a continuous LTLO$\leftrightarrow$LTT transition. The temperature range of
the LTLO phase is very sensitive to excess oxygen, and likely to shrink under more
reducing annealing conditions.

\subsection{Resistance}
\label{rho}
\begin{figure}[t]
\center{\includegraphics[width=1\columnwidth,angle=0,clip]{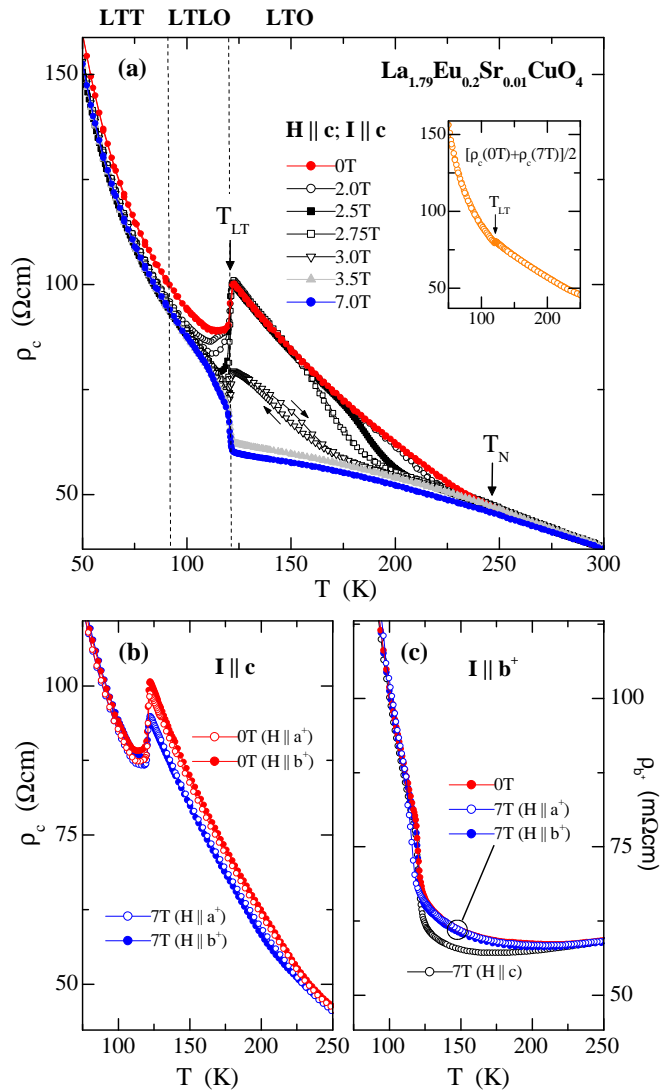}}
\caption[]{(color online) $c$-axis and $ab$-plane resistivity as a function of
temperature for different directions of the magnetic field. (a) $\rho_c$ for
$H \parallel c$. The inset shows the average of the 0~T and 7~T data sets. (b)
$\rho_c$ for $H \parallel a^+$ and $b^+$. (c) $\rho_{b^+}$ for $H \parallel a^+$,
$b^+$, and $c$.} \label{fig3}
\end{figure}

Figure~\ref{fig3}(a) shows the $c$-axis resistivity $\rho_c(T)$ for different magnetic
fields $H\parallel c$. The overall trend is an insulating behavior. However, the magnetic
field dependence reveals some dramatic changes as a function of temperature. Above the
Neel temperature of $T_N=248$~K the field dependence is very small. Between $T_N$ and
$T_{\rm LT}$ a strong decrease of $\rho_c$ with increasing $H$ is observed. Finally, in
the LTT phase the field dependence is again small. Right at the transition one can see
that $\rho_c(0T)$ decreases on cooling, while $\rho_c(7T)$ increases by an equal amount.
This shows that the $c$-axis transport in the LTT phase is distinct from both the zero
field and the high field regime in the LTO phase.
Interestingly, the average $[\rho_c(0T) + \rho_c(7T)]/2$ shows no significant change at
$T_{\rm LT}$ suggesting that primarily the magnetic scattering dependent transport is
affected by the structural transformation.

\begin{figure}[t]
\center{\includegraphics[width=0.95\columnwidth,angle=0,clip]{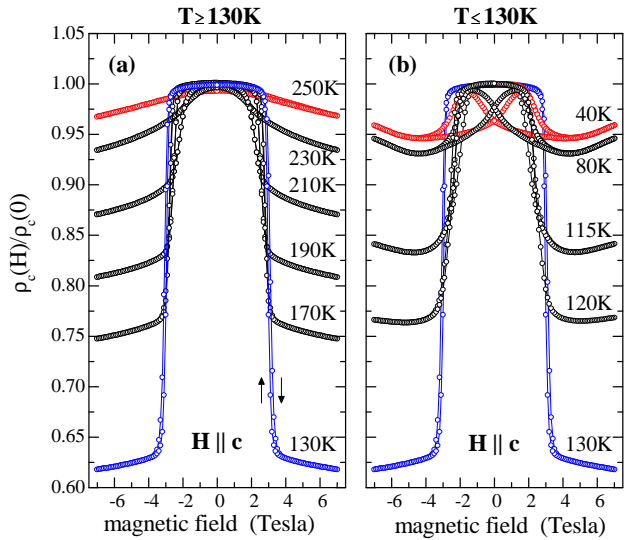}}
\caption[]{(color online) Magneto-resistance $\rho_c(H)$ at different temperatures. (a) In the LTO
phase. (b) In the LTLO and LTT phases and at 130~K.} \label{fig4}
\end{figure}

The nature of the changes $\rho_c$ undergoes at the structural transition for $H
\parallel c$ is even more obvious in the magneto-resistance curves in Fig.~\ref{fig4}. In
the AF ordered LTO phase $\rho_c(H)$ shows a sharp drop which grows with decreasing
temperature and reaches $\sim 35\%$ at 130~K, Fig.~\ref{fig4}(a). This is so to speak the
normal behavior that is also observed in the AF ordered LTO phase of pure \lsco
.~\cite{Thio88,Ando03a} It is well established, that the effect is connected to the
\textit{spin-flip} transition at $H_{\rm SF}$ which alters the spin structure along the
$c$-axis.~\cite{Thio88} Corresponding magnetization data for \lescoex\ will be discussed
in Sec.~\ref{mag}. The new observation is that, in the LTLO and LTT phases, this jump in
$\rho_c(H)$ quickly decreases, becomes hysteretic, and at $T=40$~K amounts to $\sim$5\%
only, Fig.~\ref{fig4}(b). A microscopic interpretation is given in Sec.~\ref{discussion}.

\begin{figure}[t]
\center{\includegraphics[width=0.95\columnwidth,angle=0,clip]{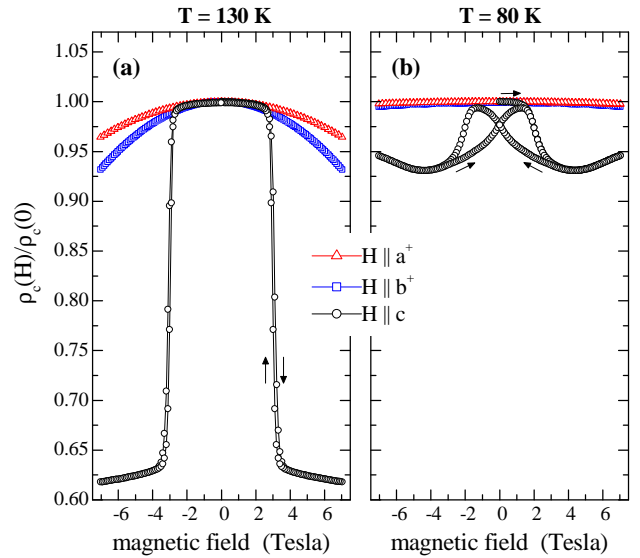}}
\caption[]{(color online) Magneto-resistance $\rho_c(H)$ for $H \parallel a^+$, $b^+$, and $c$. (a) In
the LTO phase at $T=130$~K. (b) In the LTT phase at $T=80$~K.} \label{fig5}
\end{figure}

A much weaker field dependence of $\rho_c$ was observed for $H\parallel a^+$ and
$H\parallel b^+$. Note that the crystal used for the $\rho_c$ measurements is largely
detwinned, i.e., for 80\% of the sample $b^+\parallel b$. Figure~\ref{fig3}(b) compares
$\rho_c(T)$ for $H=0$~T and 7~T. Figure~\ref{fig5} compares $\rho_c(H)$ for all three
field directions at $T=130$~K in the LTO phase and at $T=80$~K in the LTT phase. In the
LTO phase a negative magneto-resistance of several percent at 7~T is observed, which is
slightly larger for $H
\parallel b^+$ than for $H \parallel a^+$, consistent with results for \lsoco
.~\cite{Ando03a} In the LTT phase this weak magneto-resistance decreases by one order of
magnitude. It is well known that in \lco\ and in \lsoco\ a \textit{spin-flop} with
concomitant features in the magneto-resistance occurs for $H\parallel b$ and critical
fields up to 20~T, depending on the temperature.~\cite{Thio90,Ando03a,Ono04a} In
Ref.~\onlinecite{Huecker04b} it was suggested that the spin-flop field may decrease
substantially in the LTT phase. Based on the current data one can safely say that at
least up to 7~T no spin-flop takes place in the LTT phase of \lescoex .

Measurements of the in-plane resistivity $\rho_{b^+}$ are presented in
Fig.~\ref{fig3}(c). The crystal used here is only slightly detwinned, i.e., for 55\% of
the sample $b^+
\parallel b$. At zero field $\rho_{b^+}(T)$ shows a minimum at 200~K and a sharp increase at
the LTO$\rightarrow$LTLO transition. For $H\parallel a^+$ and $H\parallel b^+$ the
magneto-resistance at 7~T is very small and barely visible in the $T$-dependent data.
Field loops $\rho_{b^+}(H)$ at fixed temperature show a negative magneto-resistance of
less than 1\% at 7~T in the LTO phase and a one order of magnitude smaller effect in the
LTT phase (not shown).

For $H\parallel c$ a significant decrease of $\rho_{b^+}(T)$ is observed in the AF
ordered LTO phase, reaching 8\% at 130~K and 7~T, see Fig.~\ref{fig3}(c). Furthermore,
the field loops $\rho_{b^+}(H)$ show the same type of sharp drop at $H_{\rm SF}$ as for
$\rho_{c}(H)$ and $H\parallel c$, just much smaller (not shown). In the LTT phase the
magneto-resistance is again extremely small. Intuitively it is not obvious why, in the
LTO phase, the in-plane resistivity should decrease at a transition that effects how the
spin sublattices are staggered along the $c$-axis, but leaves the in-plane spin structure
unchanged. The first explanation that comes to mind is that, because of the extreme
anisotropy $\rho_c/\rho_{\rm ab} \sim 10^3$, a minor misalignment of the crystal or of
the contacts caused an admixture of a $c$-axis component. Since the crystal for
$\rho_{b^+}$ was quite small we cannot rule out this source of error. On the other hand,
similar observations have been reported for the LTO phase of \lsco
.~\cite{Lacerda94,Ando03a} In recent theoretical studies the effect was ascribed to a
less anisotropic localization length in the high field regime ($H>H_{\rm SF}$)
.~\cite{Shekhtman94a,Kotov07a} It was suggested that this results in a more 3D like
variable-range-hopping, making more out-of-plane states available for $ab$-plane
transport. Assuming this is true, it is clear from the present data that this channel
and, thus, $\rho_{b^+}$ become independent of $H\parallel c$ in the LTT phase, because as
the spin-flip induced magneto-resistance of $\rho_c$ disappears, so does the associated
change of the $c$-axis localization length.

\subsection{Magnetization}
\label{mag}
\begin{figure}[t]
\center{\includegraphics[width=0.95\columnwidth,angle=0,clip]{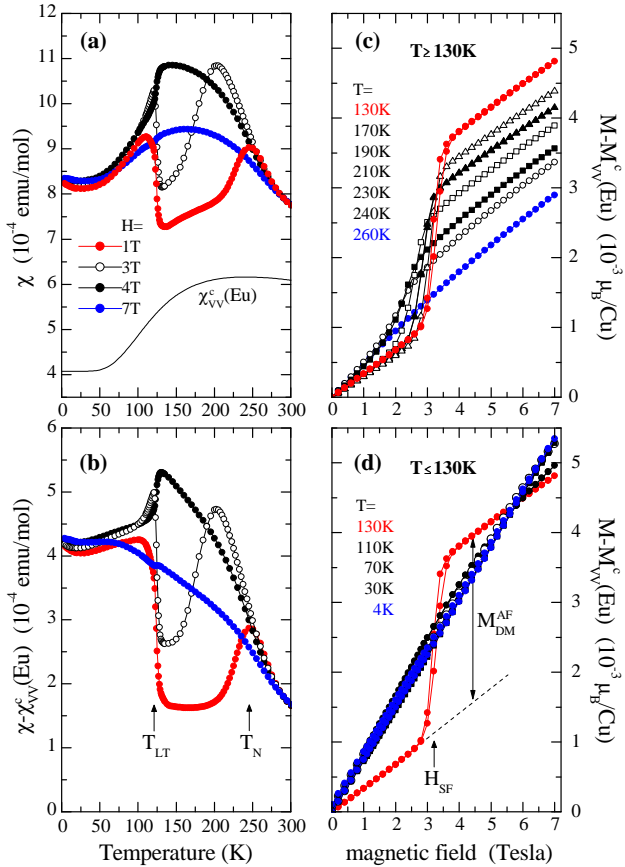}}
\caption[]{(color online) (a) Static susceptibility $\chi(T)$  for different $H \parallel c$.
(b) After subtraction of the Van Vleck contribution $\chi^c_{\rm VV}(\rm Eu)$ of
the europium ions. Right: Magnetization $M(H)$ after subtraction of the linear
europium contribution $M^c_{\rm VV}(\rm Eu)$.
(c) In the LTO phase. (d) In the LTLO and LTT phases and at 130~K.} \label{fig6}
\end{figure}
The magnetization measurements were performed on a bulky $m=0.6$~g single crystal. Note
that similar measurements on a \leco\ crystal and on \lesco\ polycrystals have been
discussed in Ref.~\onlinecite{Huecker04b}. The present sample is our first lightly doped
crystal and features very sharp transitions. The presentation of data will be limited to
$H
\parallel c$, since no significant effects have been observed for $H \parallel ab$ and
fields up to 7~T, consistent with the absence of significant magnetic field effects in
$\rho_{b^+}$.

Figure~\ref{fig6}(a) presents the static \sus\ $\chi(T)$ for different $H
\parallel c$.
\begin{figure}[t]
\center{\includegraphics[width=0.8\columnwidth,angle=0,clip]{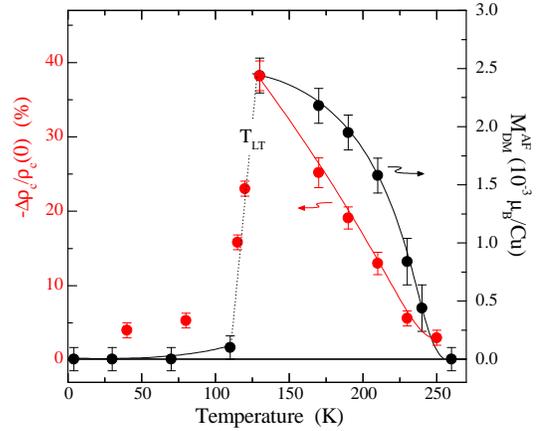}} \caption[]{(color
online) Resistivity drop $\Delta\rho_c = \rho_c(7T) - \rho_c(0T)$ in percent of
$\rho_c(0T)$ and magnetization jump $M_{\rm DM}^{\rm AF}$ at the spin-flip transition.
See Fig.~\ref{fig6}(d) for definition of $M_{\rm DM}^{\rm AF}$. Note that $M_{\rm
DM}^{\rm AF}$ is only the AF ordered part of the WFM. This part becomes zero in the LTT
phase, whereas the total size of the WFM does not change at the
transition.~\cite{Huecker04b}} \label{fig7}
\end{figure}
The Van Vleck \sus\ $\chi^c_{\rm VV}(\rm Eu^{\rm 3+})$ of the europium ions provides by
far the largest contribution (solid line). Figure~\ref{fig6}(b) shows the same data after
subtraction of $\chi^c_{\rm VV}(\rm Eu^{\rm 3+})$, which can now be compared to pure
\lsco . For $H=1$~T a sharp N\'eel peak at $T_N$ and a jump at $T_{\rm LT}$ are observed.
For $H=3$~T and higher fields the \sus\ in the AF ordered LTO phase starts to increase
significantly. The same behavior is observed in \lco .~\cite{Cheong89,Huecker04b} In
contrast, in the LTLO and LTT phases the \sus\ is elevated at any field and shows almost
no field dependence. At $H=7$~T the \sus\ increases monotonous with decreasing $T$.

As is well documented, the N\'eel peak is the fingerprint of a weak spin canting
perpendicular to the \pla\ planes, caused by \DM\ (DM)
superexchange.~\cite{Thio88,Thio94,Tabunshchyk05a,Benfatto06a} Each plane carries a weak
ferromagnetic moment (WFM) which orders antiparallel in adjacent layers for $T<T_N$. When
the external field $H\parallel c$ exceeds the spin-flip field $H_{\rm SF}$, needed to
overcome the interlayer coupling $J_\perp$, the spin lattice of every other layer rotates
by $\theta = 180^\circ$, with the effect that the WFM of all planes become parallel to
the field. As a result the \sus\ in the LTO phase increases. Note that this is the reason
why for $H = 3$~T the peak does not represent $T_N$, but the temperature below which
$H_{\rm SF}>H$ and WFM start to order antiparallel.

The changes across the LTO$\leftrightarrow$LTLO$\leftrightarrow$LTT transition are also
apparent in the magnetization curves $M(H)$. The data in Figs.~\ref{fig6}(c) and
~\ref{fig6}(d) are after subtraction of the linear $\rm Eu^{3+}$ Van Vleck contribution.
In the LTO phase the spin-flip transition grows sharper and larger for $T<T_N$. Again,
this is the normal behavior found in \lsco .~\cite{Thio94,Ando03a} In contrast, below the
structural transition no spin-flip transition is observed. The $M(H)$ curves are close to
being linear in the studied field range, indicating a significant change of the magnetic
coupling between the planes. Close to $T_{\rm LT}$ the magnetization at maximum field in
the LTO and LTT phase differs only slightly. The \sus\ at 7~T in Fig.~\ref{fig6}(b) shows
even better that there is no significant anomaly at the
LTO$\leftrightarrow$LTLO$\leftrightarrow$LTT transition. This implies that the WFM do not
change their size across the transition, and at 7~Tesla are ferromagnetically aligned in
all three phases.
There is a small number of interesting theoretical studies on this new magnetic state,
motivated by experiments on \lndco .~\cite{Viertio94a,Bonesteel93b} However, the static
magnetization presented in Fig.~\ref{fig6} and in Ref.~\onlinecite{Huecker04b} seems to
escape these earlier calculations, in particular with respect to the structure dependence
of the $M(H)$ curves and the saturation field and moment of the WFM in the LTT
phase.~\cite{MAGLTT}

Figure~\ref{fig7} compares the resistivity drop $\Delta \rho_c = \rho_c(7T)-\rho_c(0T)$
with the moment change $M_{\rm DM}^{\rm AF}$ at the spin-flip transition. In the LTO
phase the data are qualitatively the same as for \lsco ,~\cite{Thio88,Ando03a} whereas in
the LTLO and LTT phase one can see the dramatic drop of these quantities. Note that
$M_{\rm DM}^{\rm AF}$ reflects the AF coupled part of the WFM only. The total WFM, which
also consists of a non-spin-flip part (in particular in the LTT phase), continues to grow
on cooling (cf. Fig.~22 in Ref.~\onlinecite{Huecker04b}).

\section{Discussion and Conclusions} \label{discussion}
\label{dis}

\begin{figure}[b]
\center{\includegraphics[width=0.9\columnwidth,angle=0,clip]{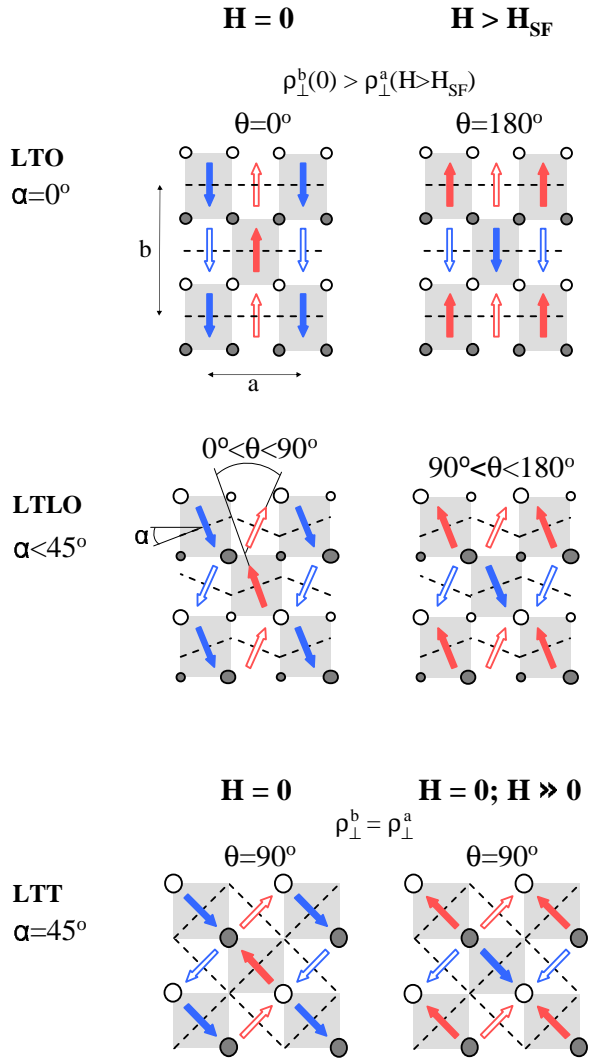}} \caption[]{(color
online) Spin structures of two adjacent $\rm CuO_2$ planes in the LTO, LTLO and LTT
phases for zero and high magnetic fields $H \parallel c$. Closed spins and gray
plaquettes form one plane, open spins and white plaquettes the other. White (gray) oxygen
atoms are displaced above (below) the $\rm CuO_2$ plane. The size of the circles grows
with displacement. Dashed lines indicate the octahedral tilt axes. Spin canting is
coupled to the octahedral buckling pattern, although canting angles ($\lesssim
0.2^\circ$) are much smaller than tilt angles ($\lesssim 5^\circ$). Spins pointing
towards white (gray) oxygen atoms are canted out of the plane (into the plane) of the
paper. $\rho_\perp^a$ and $\rho_\perp^b$ symbolically denote the microscopic (not
macroscopically measured) interlayer hopping resistance in $a$ and $b$ direction.}
\label{fig8}
\end{figure}

In several theoretical studies, motivated by the experiments on \lco\ and \lsoco , it was
pointed out that the electronic transport between the \pla\ planes does not depend on the
direction of the weak ferromagnetic moments, but on the relative orientation $\theta$ of
the spin $S=1/2$ sublattices in neighbor planes.~\cite{Thio90,Shekhtman94a,Kotov07a} The
apparent reason is that holes in an antiferromagnet prefer to hop between sublattices
with same spin direction. As is shown schematically in Fig.~\ref{fig8} for the LTO phase,
this implies that interlayer hopping at low fields ($\theta = 0^\circ$) takes place
predominantly along the $b$-axis, whereas above the spin-flip field ($\theta =
180^\circ$) it takes place predominantly along the $a$-axis. The negative $c$-axis
magneto-resistance then requires that, microscopically (not measured), the interlayer
hopping resistance along $a$ in the high-field regime is smaller than along $b$ in the
low-field regime ($\rho_\perp^a < \rho_\perp^b$). An intuitive explanation for this is
that $a<b$, although the details are known to be more
complicated.~\cite{Shekhtman94a,Kotov07a}

In the LTT phase the situation is quite different (Fig.~\ref{fig8}). The octahedral tilt
axes have rotated by $\alpha = \pm 45^\circ$ in adjacent layers. The magnetization
measurements on \lesco\ presented here and in Ref.~\onlinecite{Huecker04b}, as well as
neutron diffraction experiments on \lnco\ in Ref.~\onlinecite{Keimer93} show that, due to
DM superexchange, spins follow the alternating rotation of the tilt axes. This means that
spins are canted out-of-plane, but now form a non-collinear spin structure. Both the
tetragonal symmetry ($a=b$) and the non-collinear spin structure ($\theta = 90^\circ$)
cause a frustration of the interlayer superexchange, resulting in the absence of a well-defined
spin-flip in the $M(H)$ curves, see Fig.~\ref{fig6}(d). Moreover, the two sketched LTT
spin configurations with antiparallel (left) and parallel (right) alignment of the WFM
are energetically nearly equivalent, and should both populate the ground
state.~\cite{MAGLTT}

What are the consequences for the $c$-axis magneto-transport in the LTT phase? Because
$a=b$ and $\theta = 90^\circ$, both interlayer hopping directions are structurally and
magnetically equivalent. Moreover, the application of a high magnetic field $H||c$ has no
effect on $\theta$, although it shifts the magnetic ground state towards the one in the
right panel with parallel WFM. Hence, the LTT phase is expected to be "spin-valve"
inactive, consistent with the dramatic decrease of the magneto-resistance in
Fig.~\ref{fig3}(a) and Fig.~\ref{fig4}(b).

The remaining magneto-resistance of $\rho_c$ for $H\parallel c$ and its field hysteresis
at low temperatures, Fig.~\ref{fig4}(b), still lack interpretation. It is unclear whether
these features are intrinsic or due to structural imperfections of the LTT phase,
resulting from a limited domain size and LTLO or LTO like domain
boundaries.~\cite{Zhu94a} Nevertheless, these features seem to correspond with the
hysteresis and remanent moment observed in the magnetization curves throughout the entire
AF ordered LTT phase of \lesco\ ($x<0.02$).~\cite{Huecker04b}

\begin{figure}[h]
\center{\includegraphics[width=0.75\columnwidth,angle=0,clip]{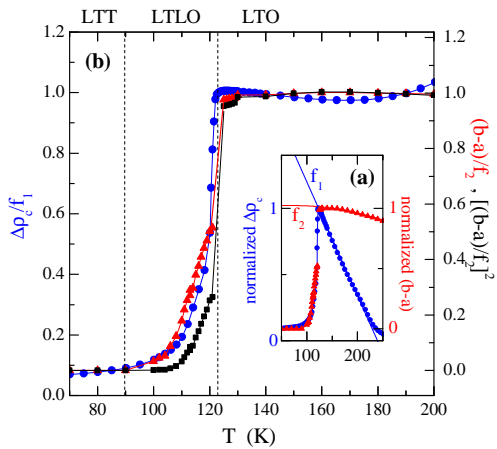}}
\caption[]{(color online) Temperature dependence of the magneto-resistance $\Delta
\rho_c$ and strain $(b-a)$ in the LTLO phase. (a) $\Delta \rho_c$ (blue circles) and
$(b-a)$ (red triangles) normalized at $T_{\rm LT}$. $f_1$ and $f_2$ are extrapolations of
the data in the LTO phase to $T<T_{\rm LT}$. (b) Same properties as in (a) divided by
$f_1$ and $f_2$. The square of $(b-a)/f_2$ is plotted as well (black squares). In the
case of the x-ray results only single phase data points are shown.} \label{fig9}
\end{figure}

The LTLO phase, represented by the middle panels in Fig.~\ref{fig8}, is expected to show
some intermediate behavior. In the temperature range $90~{\rm K} \lesssim T \lesssim
120~{\rm K}$, where this phase assumes 100\% volume fraction, it offers a unique
opportunity to study the interlayer magneto-transport as a function of $(b-a)$ and
$\theta=2\alpha$. Figure~\ref{fig9}(a) shows the temperature dependence of $(b-a)$ and
$\Delta \rho_c$ for $H \parallel c$, normalized to their values at $T_{\rm LT}$. The
correct way to compare these properties is after division by their values in the LTO
phase, extrapolated into the LTT phase; see functions $f_1$ and $f_2$. The result is
shown in Fig.~\ref{fig9}(b).
Several scenarios are possible. If $\Delta \rho_c$ (blue circles) depends primarily on
the spin orientation $\theta$, then it should be proportional to ${\rm
cos}(\theta)\propto (b-a)$ (red triangles). However, it is more likely that $\Delta
\rho_c$ also depends on the orthorhombic strain, which produces the anisotropy of the
interlayer hopping along $a$ and $b$ in first place, so that one may expect $\Delta
\rho_c$ to be in first approximation proportional to $(b-a) \cdot {\rm
cos}(\theta)\propto (b-a)^2$ (black squares).
The similarity between the temperature dependencies of $\Delta \rho_c$ and $(b-a)^q$
clearly shows that these quantities are connected.
Within the experimental error of the independent x-ray diffraction and magneto-resistance
measurements it is, however, not possible to decide on the exponent $q$. To isolate the
effects of $(b-a)$ and $\theta$ on $\Delta \rho_c$, one could study the
magneto-resistance in the LTO phase under pressure. Pressure is known to reduce the
orthorhombic strain.

What has been learned that may apply to the stripe ordered LTT phase of \lbco ? The
stripe phase consists of spin stripes coupled antiphase across the charge stripes (Fig.~3
in Ref.~\onlinecite{Huecker05a}). The stripe direction is parallel to the Cu-O-Cu bond
but rotates by $90^\circ$ in adjacent planes, similar to the octahedral tilt axes. In
zero field spins are parallel to the stripes, resulting in a non-collinear spin structure
($\theta = 90^\circ$). It is easy to see that for this type of system a large normal
state $c$-axis magneto-resistance may not be observed for any field direction. First the
tetragonal symmetry offers no advantage for any interlayer hopping direction. Second in
terms of the simple hopping picture often applied to the lightly doped compounds, i.e.,
hole and spin swap sites, interlayer hopping in the stripe phase always creates
frustrated spin moments between antiphase spin stripes. The application of a high
magnetic field $H\parallel c$ was shown to have no effect on the magnetic
order.~\cite{Huecker05a} Even if spin stripes carry a WFM due to DM superexchange (which
is still unknown) the net WFM of each plane cancels out because of the phase shift by
$\pi$ across charge stripes. Hence, no spin-flip transition can be induced, ruling out a
similarly strong $c$-axis magneto-resistance as in the AF ordered LTO phase of the
lightly doped compounds. Application of a high magnetic field parallel to the \pla\
planes produces a collinear spin structure, i.e., spins within the stripes rotate until
they are approximately perpendicular to the field.~\cite{Huecker05a} However the stripes
themselves do not rotate. Thus, even if slight lattice distortions due to, e.g., the
charge stripes would lift the structural frustration, the topology of the interlayer
superexchange for a collinear spin configuration compares to a checkerboard pattern with
$\theta=0^\circ$ and $\theta=180^\circ$. As discussed in
Refs.~\onlinecite{Li07c,Berg07a,Tranquada08a}, this indeed perfect magnetic and
electronic decoupling of the planes seems responsible for the frustration of the
interlayer Josephson coupling and the concomitant loss of 3D superconducting phase
coherence.

%
In summary, the magneto-transport of lightly hole doped \lescoex\ has been explored and
linked to structural and magnetic properties. It was shown that the low temperature
structural transition from orthorhombic to tetragonal symmetry and from collinear to
non-collinear spin structure eliminates the spin-valve type contribution to the
interlayer magneto-resistance. After calculating out spin orientation dependent effects
by averaging high and low field data, the interlayer transport appears largely unaffected
by the structural transition. In contrast, the transition triggers a significant increase
of the in-plane charge carrier localization.
\section{Acknowledgement}
The author thanks J. M. Tranquada for fruitful discussions, and P. Reutler and G.
Dhalenne for support during the crystal growth experiment at the Laboratoire de
Physico-Chimie de l'Etat Solide in Orsay. The work at Brookhaven was supported by the
Office of Science, U.S. Department of Energy under Contract No.\ DE-AC02-98CH10886.
%

\begin{thebibliography}{10}
\bibitem{HTT}
In \lsoco\ this transition occurs at $T_{\rm HT}\simeq 500$~K. For \lescoex\ we
  estimate that $T_{\rm HT}\simeq 680$~K, based on our experience that $T_{\rm
  HT}$ increases approximately 9~K per 0.01 Eu.
\bibitem{MAGLTT}
The $M(H)$ curves in the LTT phase are still poorly understood. There is no sharp
spin-flip, as well as no spontaneous weak ferromagnetism ($H_{\rm SF}\sim 0$~T). Although
in the low field regime $dM/dH$ is significantly larger than in the LTO phase, which
indicates that the AF interlayer coupling is much weaker, it still takes very high fields
($> 5$~T) to ferromagnetically align all WFM. Let us describe the AF order of the LTO
phase as $udududududududududud$, where $d$ and $u$ means WFM moment up and down. Then,
the zero field ground state of the LTT phase may possibly look like
$uudduddddduuudddudud$. There are random stacks with $udud$, $uuu$ and $ddd$ of different
length. The size of these domains is likely to vary also in the $ab$ plane, and will most
likely be smaller than in the LTO phase (due to the first order nature of the
LTO$\leftrightarrow$LTLO transition). This may cause a broad distribution of local
critical fields and, thus, smear out the spin-flip to a degree that the $M(H)$ curves
become effectively linear in a wide field range.
\end{thebibliography}
%

\end{document}